\documentclass[12pt]{article}
\usepackage{graphicx}
\usepackage{cite}
\usepackage{amsmath}

\begin{document}
\newcommand {\be}{\begin{equation}}
\newcommand {\ee}{\end{equation}}
\newcommand{\bea}{\begin{eqnarray}}
\newcommand{\eea}{\end{eqnarray}}
\newcommand{\ba}{\begin{array}}
\newcommand{\ea}{\end{array}}

\renewcommand{\th}{{\rm tanh}}
\newcommand{\sech}{{\rm sech}}
\newcommand{\F}{{\cal F}}
\newcommand{\bfDel}{\mbox{\boldmath $\Delta$}}
\newcommand{\bfwp}{\mbox{\boldmath $\wp$}}
\newcommand{\bfro}{\mbox{\boldmath $\rho$}}
\newcommand{\bfd}{\mbox{\boldmath $\delta$}}
\newcommand{\bfxi}{\mbox{\boldmath $\xi$}}
\def\max{\mathop{\rm max}}
\def\Sup{\mathop{\rm Sup}}
\def\m{{\rm m}}
\def\M{{\rm M}}
\def\ch{{\rm ch}}
\def\l{{\rm l}}
\def\rr{{\bf r}}
\def\RR{{\bf R}}
\def\R{{\rm R}}
\def\k{{\rm k}}
\def\E{{\bf E}}

\title{Mechanism of Threshold Elongation of DNA  Macromolecule}
\author{Sergey N. Volkov\\
Bogolyubov Institute for Theoretical Physics, NAS of
Ukraine,\\14-b Metrolohichna Str., Kyiv, 03680, Ukraine \\
snvolkov@bitp.kiev.ua }\maketitle

\setcounter{page}{1}%
\maketitle

\begin{abstract}
The mechanism of threshold elongation (overstretching) of DNA macromolecules under the action 
of external force is studied within the framework of phenomenological approach. When considering 
the task it is taken into account that double-stranded DNA is a polymorphic macromolecule with a 
set of metastable states. Accordingly to proposed mechanism, DNA threshold elongation can take 
place as a cooperative structural transition of macromolecule to metastable form, which is 
stabilized by the external force. For the description of DNA overstretching, the model included 
external (stretching) and internal (conformation) displacement components is constructed. As 
assumed, both components are coupled on the pathway of double helix transformations. It is shown, 
that under force action DNA deformation proceeds in two stages. First one is the restructuring 
of the double helix, which leads to formation of conformational bistability 
in DNA macromolecule. On the second stage, the conformational transition and the deformation
induced by it consistently cover the macromolecule as a threshold process. The propagation of
overstretched deformation in bistable macromolecule occurs as the motion of transition
boundaries (domain walls), which are topologically stable structural excitations.
Using the proposed approach, the contributions to DNA overstretching process are determined.
This is elastic stretching of the double helix, deformation induced by conformational
changes in macromolecule before overstretching transition, and threshold elongation as such.
The greatest contribution to overstretching process in DNA (more then 80$\%$) is introduced
by threshold elongation. As shown, the effect of threshold deformation in DNA double helix
takes place due to the critical correspondence of acting force with elastic properties of
macromolecule, and achieves by restructuring of the double helix. The comparison of calculated
characteristics of overstretching process for heteronomous DNA with experimental data shows good
agreement. Analysis of the results obtained, together with the available literature data allow
to conclude, that overstretching transition in the double helix is dynamical process and can
spread in DNA chain. In the same time, in DNA with A$\cdot$T-rich content due to large 
dissipation the overstretching process leads to force induced melting transition, and should
have nearly static character.
\end{abstract}

\section{Introduction}\label{sec1}

Mechanical properties of DNA macromolecule are essentially important for understanding the
mechanisms of biological processes in cells. To interpret the experimental results on DNA
mechanics, the macromolecule has been often modeled as an elastic rod which parameters
determined in various indirect experiments \cite{Schellman74,BarkleyZimm79,F-K83}.
Due to development of single molecule manipulation techniques, the mechanical parameters
of DNA double helix (Young's module, bending and rotating stiffness) have been significantly
refined \cite{SSmith+B92,Cluzel+96,SSmith+B96,Bustamante00,Lavery02,Bustamante03}.
A new tool has been also used to study such important mechanical processes as DNA unzipping,
stretching, unwrapping, packaging, looping and interaction with proteins
\cite{Bustamante00,Lavery02,Bustamante03,Bockelmann04,Allemand06,Prevost09,PhysLife-Williams10,Killian12}.

\begin{figure}[h!]
\begin{center}
\includegraphics[width=8.0cm]{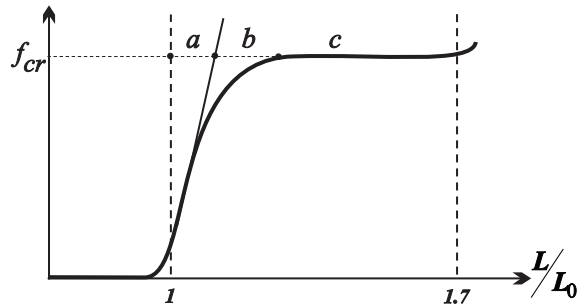}
\caption{Typical curve for dsDNA stratching. The figure is drawn in accordance with data
\cite{SSmith+B96}, where the overstretching of $\lambda$-phage DNA is studied ($f_{cr}$=65 pN).
The contributions to DNA overstretching deformation are discriminated in the present paper as:
($a$) - elastic stretching, ($b$) - deviation from elastic stretching, ($c$) - overstretching
transition as such.}
\label{overstretch}
\end{center}
\end{figure}

New results of DNA mechanics satisfy the elastic rod model not always. Sometimes, under the
action of external force the deformation of double-stranded (ds) DNA occurs cooperatively as
a threshold process (see reviews \cite{Bustamante00,Lavery02,Bustamante03,Bockelmann04}).
Most clearly such effects manifest themselves in the experiments on DNA chain stretching.
It is shown that at some critical value of applied force ($f_{cr}$$\sim$ 65 pN)
the rotationally unconstrained macromolecule overstretches, that is, elongates in 1.5-1.7
times \cite{Cluzel+96,SSmith+B96,Bustamante00,Lavery02,Bustamante03}. The constrained
dsDNA transforms to overstretched state under the action of larger force ($\sim$ 110 pN)
\cite{Mameren09,PaikPerkins11}. The typical view of dsDNA overstretching curve drawn by data
\cite{SSmith+B96} is shown in Fig. \ref{overstretch}.

Similar results are observed for synthetic polynucleotides \cite{Rief99,Gaub2000}
and oligonucleotides \cite{FuMarcoYan10,FuYan11,Norden12}. These
macromolecules have a DNA-type structure with two bonded polynucleotide strands.
The stretching of single-stranded DNA corresponds to behaviour of extensible elastic rod
\cite{Bustamante00,Bustamante03}. For double-stranded RNA macromolecules the effect
of overstretching is not observed \cite{Liphard2001}.

A plenty of works focusing on the mechanism of DNA overstretching has been performed.
As considered \cite{SSmith+B92,SSmith+B96,Bustamante00,Lavery02,Bustamante03}, the
force-extension curve (Fig. \ref{overstretch}) on the interval of small forces ($<$6 pN)
describes the process of dsDNA unwrapping, for higher forces - the double helix elastic
stretching. Then, on the interval before threshold elongation, the deviation
from elastic trajectory of DNA stretching curve is observed, and at critical force value
the threshold elongation as such occurs. The nature of the first two contributions
to dsDNA elongation has been well studied, but the origin of last two processes remain
not clear still.

Basically, there are two scenarios of the process. First one is force-induced internal melting
(FIM) of DNA double helix with a gradual slow transition into ss-state \cite{PhysLife-Williams10,Rouzina2001a,Rouzina2001b,Mameren09,Wuite11}. Second scenario is
cooperative fast transformation from the usual dsDNA ($B$-form) to the stretched form ($S$)
with preserved hydrogen bonds in complementary pairs
\cite{Cluzel+96,SSmith+B96,Rief99,Danilovich09,Maaloum11,Bianco11}. This is so called
$B$-$S$ transition. Thermodynamical analysis made in \cite{CoccoMarco04,Whitelam08} and experiments
\cite{FuMarcoYan10,FuYan11,Bianco11,Norden12,ZhangYan12} show that both FIM and $B$-$S$ transitions
are possible for dsDNA. Their realization depends on temperature, counterions concentration in
solution, and DNA nucleotide content. Studying the kinetics of DNA overstretching authors of
\cite{FuYan11,Bianco11} have come to the conclusion that overstretching process
takes place initially as $B$-$S$ transition, and then FIM transition can occur.

It is also shown, that threshold elongation of dsDNA is no uniform deformation
\cite{Mameren09,PaikPerkins11,Maaloum11}, and the fragments with different states
(stretched and overstretched) can alternate in one chain under critical force action.

Recently, one more step in understanding the mechanism of DNA overstretching transition
has been done. In paper \cite{Norden12} the appearance of bistability between $B$ and $S$
forms of dsDNA is observed in the force interval, when overstretching is occurred in dsDNA.
The similar event is observed  under study of dsDNA stretching through the melting
of one DNA strand from another - unpeeling \cite{Wuite11}(Suppl.).

As it turns out, the course of overstretching process in dsDNA is recognized
in general terms. However, this is not the case for the mechanism.

The conformational analysis and molecular dynamics modeling have found few structures of
overstretched DNA depending on the value of applied force and the manner of its applying
to macromolecule \cite{Cluzel+96,LebrunLavery96,KonradBlonick96,Lauton05,Prevost09}.
The most probable structure of $S$-DNA is a narrow fibre with inclined hydrogen bonded base
pairs and saved stacking interactions. This structure is obtained in conformational analysis
studies \cite{Cluzel+96,LebrunLavery96}, and observed in MD simulations of DNA elongation
in solvent conditions \cite{KonradBlonick96,Lauton05}. The performed calculations demonstrate
the opportunities of the double helix to overstretching transitions, but no offer an
explanation of their appearance in the double helix.

The mechanism of DNA overstretching is proposed in \cite{Bianco11}, where authors believe
that overstretching deformation in dsDNA happens as two-state reaction process - transition
from ground compact to metastable extended state, within the appearance of some domains in
macromolecule chain. But authors \cite{Bianco11} do not explain, why the monomers in the
domains remain in unprofitable metastable states, and do not return to usual compact form.

Authors of \cite{Marko97,Gore06,StormNelson} have suggested that the coupling of DNA stretching
with double helix twisting and bending can be the reason of DNA threshold elongation. It should be
noted, that of course, the coupling of deformation components of macromolecule, should give the
definite quantitative effect in dsDNA stretching, but cannot be the origin of threshold process.

On the other hand, the threshold effect may be result of the coupling of macromolecule
deformation components with the conformational restructuring of the double helix. This
possibility is indicated under dsDNA overstretching modeling in \cite{LebrunLavery96},
where sharp jumps in macromolecule stretching energy are observed.
A similar concept has expressed by authors of \cite{Bianco11}. Studding the
kinetics of overstretching process, they come to conclusion, that dsDNA macromolecule
shows two kinds of elasticity in the ranges of acting force lower and higher, then 35 pN.
That is, under tension of external force the macromolecule can change its structural
organization.

So, the nature of DNA threshold elongation is still being discussed.
It is necessary to understand why and how the transition to overstretched form
occurs in DNA chain. What the action produces by external force, and how is it connected
with the parameters of the double helix? What is the conditions required for threshold
deformation of the double helix, and what is the differences between $B$-$S$ and FIM
transition realization in dsDNA?

To answer these and some other questions in present paper a phenomenological model for
DNA deformation study is proposed (\ref{sec2}). Under model construction, dsDNA
is considered as a polymorphic macromolecule, which has a set of metastable states.
Following this thesis, the contributions to deformation of polymorphic
macromolecule are determined in \ref{sec3}. The restructuring in polymorphic
macromolecule under action of external force, and the possibility of conformational
bistability formation are studied in \ref{sec4}. In \ref{sec5} the dynamics of
of conformational transitions from compact to stretched form in dsDNA is under
consideration. Using the developed model, the quantitative characteristics of dsDNA
overstretching are estimated in \ref{sec6}. The agreement of developed theory with
experiments, the dependence of DNA overstretching on nucleotide content, the reasons
for of $B$-$S$ or FIM transition realization in DNA chain are discussed in (\ref{sec7}).
At the end, the conclusions and predictions from the performed study are formulated \ref{sec8}.

\section{DNA macromolecule under the action of external force}\label{sec2}

Considering the action of external force on dsDNA, it should be taken into account
that DNA double helix is a polymorphic molecule \cite{Saenger,Blackburn06}. As in
all polymorphic structures, the form of dsDNA and its changes are coupled with the
disposition and displacements of macromolecule structural elements in the frame of
double helix. This is well illustrated by example of $B$-$A$ conformational transition
in DNA, where the parameters of the double helix turn (such as length and twist angle)
change together with the changes in positions of structural elements inside the
double helix \cite{Ivanov94,Saenger}. The interrelation between the displacements
of DNA structural elements with the changes of the double helix form manifests
itself also in the observations of conformationally induced deformations in dsDNA
\cite{Crothers,Olson00,Dickerson01}. The importance of  coupling between the double helix
deformation and its conformational state is evident also from the results of conformational
analysis of DNA overstretching in \cite{Cluzel+96,LebrunLavery96,KonradBlonick96,Lauton05}.
So, the coupling of macromolecule deformation with its conformation state will be accounted
under model construction of DNA overstretching process in the present work.

For understanding the mechanism of dsDNA overstretching deformation the model with
two types of displacement components will be used. One component will describe an external
deformation of the macromolecule, and another component - internal conformational
transformation in the double helix structure. Let external component ($R_n$) will
describe a displacement of $n$-th monomer link from its equilibrium position in
macromolecule chain under dsDNA stretching. Note, this is one of the components
of DNA model as elastic rod, for which the value of elastic constant is well known.
Let internal component will describe the change in conformation state of double helix
monomer link ($r_n$). It is well known, that one of the dominant components in specifying
the conformation of the double helix is the position of base pair \cite{Ivanov94,Saenger}.
The displacement of paired DNA bases in monomer unit reflects the changes in the double
helix conformation under overstretching transition as well, as testified by the results of \cite{Cluzel+96,LebrunLavery96,KonradBlonick96,Lauton05}. So,
the displacement of mass center of DNA base pairs with respect to mass center of monomer
unit will be considered as internal component under modeling the overstretching process
in dsDNA. In the present paper it will be studied also the dynamics of deformation in
DNA macromolecule. So, let both model components depend on time: $R_{n}(t)$ and $r_{n}(t)$.

The expression for energy of DNA macromolecule under the action of external force
can be written in two-component approach as follows:
\bea \begin{split}
E & = {1\over2} \sum_n \biggl\{M\dot{R^2}_{n} + m\dot{r^2}_{n}
+ k_{R} \bigl[R_{n} - R_{n-1}\bigr]^2 \\
&+ k_{r}\bigl[r_{n}- r_{n-1}\bigr]^2 + \Phi (r_{n}) \\
&- \chi\, F (r_{n}) \bigl[(R_{n+1}-R_{n})+
(R_{n}-R_{n-1})\bigr] \biggr\} + A(R) \, .
\end{split}
\label{eq:Energy} \eea
In expression (\ref{eq:Energy}) $\dot{R}$, $\dot{r}$ are derivatives on time; $k_{R}$ and
$k_{r}$ are the elastic constants, which describe the interactions along macromolecular
chain for external and internal components; $M$ means monomer mass, $m$ is
reduced mass of monomer link according to the pathway of conformational transformation.
It is significant to note, that for DNA conformational pathways with joint motions of
nucleic bases, the masses $M$ and $m$ have the same value for A$\cdot$T and G$\cdot$C
pairs \cite{TheorBiol90,BiophBull03,BiolPhys05}.

\begin{figure}[h!]
\begin{center}
\includegraphics[width=6.0cm]{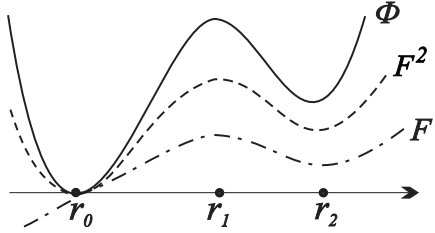}
\caption{The shapes of potential functions used in the model (\ref{eq:Energy}).}
\label{energy}
\end{center}
\end{figure}

Potential function $\Phi(r)$ describes the change of monomer energy on the pathway of
conformational transformation from the usual compact to the stretched helix form.
Under conditions when compact DNA form is more stable (at the physiological
conditions - $B$-form \cite{Saenger}), function $\Phi(r)$ must be taken in the shape of
double well with non-equivalent stable states: ground stable state - $B$, and metastable
one - $S$ (Fig. \ref{energy}). It is assumed, as in \cite{SSmith+B96,KonradBlonick96},
that potential barrier in $\Phi(r)$ under DNA overstretching transition originates from the
conformational barriers in sugar-phosphate backbone of the double helix, and mainly related
to the conformation of sugar rings.

The term with coefficient $\chi$ describes the coupling between internal and external
components under macromolecule restructuring. The value of this coupling depends on
macromolecule conformational state, determined by function $F(r)$, and on value of
external force controlled by parameter $\chi$. The coupling between internal
and external components exists and without action of external forces, as seen from
many examples in \cite{Saenger,Blackburn06}. The presence of this term and its sign
in expression (\ref{eq:Energy}) reflects the fact, that polymorphic macromolecule
can reduce the energy needed for deformation due to the change of its conformation.

Potential functions $\Phi(r)$ and $F(r)$ due to their physical meaning should increase
on the pathway from ground state of the system up to transition barrier, and then decrease
near metastable state (Fig. \ref{energy}). Explicit view of used potential functions can
be founded with the help of conformational analysis, but for the goals of
present study it is sufficient to determine the shapes of potential functions only.

In expression (\ref{eq:Energy}) the last term describes the action
of external force $f$ on the macromolecule state. It is reasonable to consider,
that the force is applied to the external component of macromolecule deformation:
\bea
A(R)=- f \sum_n (R_{n} - R_{n-1}).
\label{eq:Work}\eea
When writing expression (\ref{eq:Energy}), it is assumed also that under external force action
the macromolecule chain is fully unwrapped up to its contour length before stretching.

Under modeling the process of DNA overstretching, it will be not specified the
type of transition, $B$-$S$ or FIM, at once. The differences between them
will be indicated on the basis of obtained results at the end of the study.

Note, that in this work just the coupling between one external and one internal components
of DNA transformation is under consideration. It is clear, that in such deformation as DNA
overstretching and other degrees of freedom of the double helix have to take place also.
But in the present study the simplest model with two dominated components
is considered to understand the basic mechanism of threshold deformation in the double helix.
Such two-component approach has been successfully used for the modeling of $B$-$A$ transition
in dsDNA  \cite{TheorBiol90,BiophBull03,BiolPhys05}, and under studies of the effects of DNA
localized bending and unzipping \cite{VolkovKanevska06,VolkovSolov'ov07,VolkovSolov'ov09}.
The action of external force on DNA macromolecule in last papers is considered indirectly,
as the change in external component stiffness. In the present work the external force
is seen as a direct participant in the process of dsDNA deformation.

\section{Deformation of polymorphic macromolecule}\label{sec3}

Using model (\ref{eq:Energy},\ref{eq:Work}), let us study the deformation of
polymorphic macromolecule, its dependence from the conformational state and from
the value of external force. Let us consider firstly the uniform deformation
of polymorphic macromolecule. In this case it is true:
$R_{n}-R_{n-1}\equiv \wp$, $r_{n}\equiv r$, and $\dot{R}=0$, $\dot{r}=0$.
Accordingly to considered problem, the potential energy of macromolecule monomer
link is bound to have two minima ($r_{0}, r_{2}$), and one maximum ($r_{1}$)
between them. Let one of minima ($r_{0}$) be a ground state, and for this state
the following condition takes place: $\Phi(r_0)=F(r_0)=0$. For other extremum
points ($r_1, r_2$) let us assume $\Phi(r_1)>\Phi(r_2)>0$, and $F(r_1)>F(r_2)>0$
(Fig. \ref{energy}).

The expression for energy density of the
deformed macromolecule can be written in the form:
\bea \mathcal{E}(r,\wp) = \frac{1}{2} \bigl[ {k_{R}}\, {\wp}^2
+ \Phi(r) - 2\chi  F(r) \wp \bigr] - f \wp \, . \label{eq:uniE}
\eea
Here $\mathcal{E}=E/N$, and $N$ is the number of monomer links in macromolecule.

In the presence of acting force equilibrium states of the system
can be determined by solution of following equations:
\bea \frac {\partial\mathcal{E}}{\partial r} = \frac {d\Phi}{d r}
-2\chi  \frac {dF}{dr} \wp = 0 \, , \label{eq:deconform}\eea
\bea \frac{\partial\mathcal{E}}{\partial\wp} = k_R\, \wp
-\chi\,F(r)- f = 0 \, . \label{eq:dedeform} \eea
From last equation, the expression for deformation of monomer link
of polymorphic macromolecule can be written as:
\bea \wp = \rho_{el}+\rho_{\chi}(r) \,. \label{eq:deform} \eea
Here
\bea \rho_{el}=\frac{1}{k_{R}}f ,\,\,\, \, \rho_{\chi}(r)=\frac{\chi}{k_{R}} F(r)
\,. \label{eq:def2} \eea

As might be expected, the deformation of polymorphic macromolecule is caused by
force action, elastic properties of the chain, and conformational state of
macromolecule. First term in expression (\ref{eq:deform}) describes elastic
deformation of macromolecule as in the model of elastic rod ($\rho_{el}$).
Second term ($\rho_{\chi}$) appears in expression
(\ref{eq:deform}) due to coupling between external and internal components
of macromolecule transformation. This term describes the conformationally induced
deformation.

Let us consider the action of external force on conformational state of polymorphic
macromolecule. After substitution of expression (\ref{eq:deform}) in equation
(\ref{eq:deconform}) and subsequent integration, the expression for conformational
energy of polymorphic macromolecule under tension of external force can be written as:
\bea \mathcal{E}(r) = {1\over2}\left[\Phi(r) - \frac{{\chi}^2 }{k_R}
F^{2}(r) \right]- \frac{\chi}{k_{R}}f F(r) +
C_r \, , \label{eq:EPh-r}\eea
where $C_r$ is the constant of integration which will be determined later.

It is clear, that the view of conformational energy (\ref{eq:EPh-r}) depends on explicit
expressions of used potential functions. However, it is quite enough to study the problem
envisaging only their shapes shown in Fig. \ref{energy}. Notice, as seen from expression
(\ref{eq:EPh-r}), the shapes of $\Phi(r)$ and $F^{2}(r)$ should be similar due to description
by these functions the different aspects of the same physical process -- the energy changes
during the transition from one conformational state to another. Thus, let us study the problem
within following approximation: $\Phi(r)=\epsilon F^{2}(r)$. Here parameter $\epsilon$ accords
with the value of energy barrier between the conformations. Such approach does not limit the
generality of the study and, as will be showed, it yields sufficiently interesting results.
Using proposed approximation, expression (\ref{eq:EPh-r}) can be written as follows:
\bea \mathcal{E}(r) = {1\over2}\epsilon_{\chi}
F^{2}(r)- \frac{\chi}{k_{R}}f F(r) +
C_r \, , \label{eq:E-r}\eea
where
\bea \epsilon_{\chi}=\epsilon - \frac{{\chi}^2}{k_R} \,. \label{eq:eps}\eea
For the double-well shape of energy (\ref{eq:E-r}) the value $\epsilon_{\chi}$ should be
positive, thus, it must be considered:
\bea \epsilon k_{R} > \chi^{2} \,.
\label{eq:epsK} \eea

Expression (\ref{eq:E-r}) clearly shows, that the action of external force can affect the
conformational state of polymorphic macromolecule due to the coupling between external
and internal components. Under force action the positions of macromolecule stable states
should change, and new extremum points of macromolecule energy ($r_{0f}$, $r_{1f}$,
and $r_{2f}$) should be found. Rewriting equation (\ref{eq:deconform}) for macromolecule
conformation energy (\ref{eq:E-r}), one can obtain the equation for the determination
of new extremum states:
\bea \left[\epsilon_{\chi} F(r)- \frac{\chi}{k_{R}}f
\right] \frac{dF}{dr}= 0 \, . \label{eq:Erderiv} \eea

Note, as seen from (\ref{eq:Erderiv}), in the absence of external force
the positions of stable states of macromolecule are determined by equation:
\bea F(r) \frac{dF}{dr}= 0 \, . \label{eq:Er0deriv} \eea
The extremum $r_{0}$ is determined by the form of potential function $F(r)$
only, from the condition $F(r_{0})$=$0$. As follows from expression (\ref{eq:Er0deriv}),
the positions of two other extrema can be found from the solution
of equation: $\frac{dF}{dr}=0$, and they are $r_{1}$ and $r_{2}$.

When $f \neq 0$, from equation (\ref{eq:Erderiv}) it follows
that new extremum $r_{0f}$ can be obtained from solution of equation:
\bea F(r_{0f})= \frac{\ae}{\chi}f \, , \label{eq:Fo}\eea
where
\bea {\ae} = \frac {{\chi}^2 } {\epsilon k_R - {\chi}^2}\,>0 \, .
\label{eq:kap} \eea
Two other solutions follow again from the equation:
$\frac{dF}{dr}=0$, and they are $r_{1f}=r_{1}$, $r_{2f}=r_{2}$. When writing
expression (\ref{eq:kap}), the inequality (\ref{eq:epsK}) is taken into account.

The first derivative of conformational energy (\ref{eq:E-r}) on $r$ for point
$r_{0f}$ is zero, as seen from (\ref{eq:Fo}), and the second derivative for
$r_{0f}$ remains positive. Thus, this point is energy minimum, a new ground state
of the system in the presence of external force. New position of ground state
can be found by solving equation (\ref{eq:Fo}) with known function $F(r)$, values of
acting force, parameters $\ae$ and $\chi$.

The change of macromolecule ground state position in the presence of external force
can be understood directly from equation (\ref{eq:Fo}). It is seen, when
$f\rightarrow0$, then $F(r_{0f})\rightarrow0$, and $r_{0f}\rightarrow r_0$. On the
other hand, if  $f>0$, then $F(r_{0f})>0$. Accordingly to the shape of function $F(r)$
(Fig. \ref{energy}), it may be so, if macromolecule ground state shifts
($r_0 \rightarrow r_{0f}$) under force action.

Another energy minimum of the system is $r_{2f}$. For this extremum point the
condition of minimum is also met, because $\frac{d\mathcal{E}}{dr}\mid_{r_{2f}}=0$,
due to $\frac{dF}{dr}\mid_{r_{2f}}=0$, and, as it can be showed, the second
derivative is positive. It is clear, that between the minima ($r_{0f}$ and
$r_{2f}$) the maximum at $r_{1f}$ is realized.

To fully determine the ground state of the system under the action of external force
let us assume $\mathcal{E}(r_{0f})=0$. Then, in accordance with equation (\ref{eq:E-r})
and expression (\ref{eq:Fo}), the constant value $C_r$ can be determined as:
\bea C_r=\frac{\chi}{2k_R} \, F(r_{0f}) \,f   \,. \label{eq:Cr}\eea

Taking into account expression (\ref{eq:Fo}), and constant value
(\ref{eq:Cr}), it is possible to write expression for conformational energy
(\ref{eq:E-r}) in the form:
\bea \mathcal{E}(r) = {1\over2} \,\epsilon_{\chi}\, \mathcal{F}^{2}(r)
 \, , \label{eq:Me} \eea
where
\bea \mathcal{F}(r)=F(r)-F(r_{0f}) \,. \label{eq:E-rF} \eea

It is seen, that at $r=r_{0f}$, $\mathcal{E}(r_{0f})=0$ in accordance with condition
accepted above for the ground state.

Using the form of function (\ref{eq:E-rF}), the deformation of macromolecule monomer
link due to its conformational transformation (\ref{eq:def2}) can be presented as:
\bea \rho_{\chi}(r)=\frac{\chi}{k_{R}}[F(r_{0f})+ \mathcal{F}(r)]=
\rho_{o\chi} + \rho_{tr}(r)  \, . \label{eq:ROchi} \eea
In expression (\ref{eq:ROchi})
\bea \rho_{o\chi}=\frac{\chi}{k_{R}}F(r_{0f})= {\ae} \rho_{el} \, .
\label{eq:rooChi} \eea
Under writing expression (\ref{eq:ROchi}), equation (\ref{eq:Fo}) and expression
(\ref{eq:def2}) are used.

The value $\rho_{o\chi}$ is the primary deformation of macromolecular structure due to the
change of macromolecular conformation before the structural transition.
One of the manifestations of primary
deformation is the shift of ground state of polymorphic macromolecule. As seen from
(\ref{eq:rooChi}) and (\ref{eq:Fo}), if $f\rightarrow0$, then $F(r_{0f})\rightarrow0$ and
$\rho_{o\chi}\rightarrow0$.

The second term in (\ref{eq:ROchi})
\bea \rho_{tr}(r)= \frac{\chi}{k_{R}}\mathcal{F}(r)\,  \label{eq:rotr} \eea
is the deformation accompanied a structural transition as such, when it occurs in
macromolecule.

By this means, the deformation of the monomer link of polymorphic macromolecule under the
action of external force consists of elastic deformation ($\rho_{el}$), and the deformations
induced in macromolecule structure before ($\rho_{o \chi}$) and during ($\rho_{tr}$) the conformational transition.

\section{Critical regime: formation of conformational bistability}\label{sec4}

Thus, polymorphic macromolecule has an additional mechanisms of deformation
connected with conformational transformations in its structure. Let us consider the action
of external force on conformational state of polymorphic macromolecule and determine the
conditions, when the deformation can have a threshold character.

\begin{figure}[h!]
\begin{center}
\includegraphics[width=6.5cm]{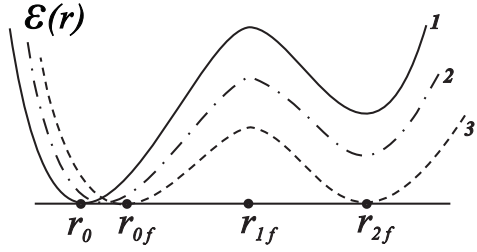}
\caption{Transformation of macromolecule conformational state under the force action:
curve 1 - $f=0$; curve 2 - $0<f<f_{cr}$ ; curve 3 - $f=f_{cr}$ (critical regime).}
\label{Crit-energy}
\end{center}
\end{figure}

As shown in previous section, the external force can shape the conformation state of polymorphic
macromolecule. One of the consequences of this process is the shift of ground state position
in macromolecule conformation. Another one can be the changes in energy barrier between
ground and metastable state, and in energy difference between conformational states as well.
Really, as seen from expression (\ref{eq:Fo}), under increasing of external force
the value of $F(r_{0f})$ should increase. So, the difference between $\mathcal{E}(r_{0f})$
and values $\mathcal{E}(r_{1f})$, and $\mathcal{E}(r_{2f})$ should become smaller,
accordingly to (\ref{eq:Me},\ref{eq:E-rF}). At some critical value of external force
($f_{cr}$) the macromolecule conformation can transit to form with two equivalent stable
states and reduced transition barrier (Fig. \ref{Crit-energy}). Such transformations in
macromolecule structure make the transitions between $r_{0f}$ and $r_{2f}$ states very probable.

The conditions of such macromolecule transformation under the action of critical force
can be found from the relation:
\bea  \mathcal{E}_{cr}(r_{0f}) = \mathcal{E}_{cr}(r_{2f}) \, .
\label{eq:e1}\eea
The realization of equality (\ref{eq:e1}) means, that conformational energy (\ref{eq:Me})
takes the form of double-well function. That is, the macromolecule transforms
into bistable conformation (see Fig. \ref{Crit-energy}, curve 3). Such a state of the
system, appeared under the external force tension, will be called as critical regime.

Taking into account expressions (\ref{eq:Me},\ref{eq:e1}), and condition
$\mathcal{E}_{cr}(r_{0f})=0$ (and hence $\mathcal{E}_{cr}(r_{2f})=0$),
it can be concluded that
\bea  \mathcal{F}_{cr}(r_{0f}) = \mathcal{F}_{cr}(r_{2f})=0 \, .
\label{eq:F1}\eea
As follows from relation (\ref{eq:E-rF}), in this case it is true:
\bea  F(r_{0f}) = F(r_{2f}) \, .
\label{eq:e11}\eea

From relations (\ref{eq:e1},\ref{eq:F1},\ref{eq:e11}) it is seen, that expression for macromolecule conformational energy in the critical regime can be approximated by the analytical function. Really,
on the interval of ($r_{0f},r_{2f}$) energy (\ref{eq:Me},\ref{eq:E-rF}) can be presented in the view:
\bea \mathcal{F}_{cr}(r) = \frac{1}{d^2}(r-r_{0f})(r_{2f}-r) \, , \label{eq:Fcr}\eea
where $d =(r_{2f}-r_{0f})/2$.

Thus, in critical regime the expression for conformational energy (\ref{eq:Me})
can be written in the form:
\be \mathcal{E}_{cr}(r) = \frac{\epsilon_{\chi}}{2d^4} \, (r-r_{0f})^{2} (r_{2f}-r)^{2} \,.
\label{eq:Ecr}\ee
As seen, function (\ref{eq:Ecr}) has two minima and one maximum ($r_{1f}=r_{0f}+d$).
Substituting  the coordinate of energy maximum in expression (\ref{eq:Ecr}),
the value of energy barrier in the critical regime can be determined as:
\be \mathcal{E}_{cr}(r_{1f}) = \frac{1}{2}\epsilon_{\chi}= \epsilon_b \,.
\label{eq:Echi}
\ee

As a whole, using expressions (\ref{eq:deform},\ref{eq:def2},\ref{eq:E-rF},\ref{eq:rooChi}),
energy density of polymorphic macromolecule
(\ref{eq:uniE}) in the conditions of critical force action can be written as:
\bea \mathcal{E}_{cr}(r,\rho) = \mathcal{E}_{cr}(r)+ \mathcal{A}_{cr}(f,\rho) \,
\, , \label{eq:uniEcr} \eea
that is, the sum of critical conformation energy (\ref{eq:Ecr}) and the work performed by
external force to form the bistable conformation of macromolecule. The value of this work
accords to the reaction of the system on external action, and per one monomer unit has the
following view:
\bea \mathcal{A}_{cr}= \frac{1}{2}\rho_{el}f_{cr}+ \frac{1}{2}\rho_{o\chi}f_{cr}
 \, , \label{eq:Acr} \eea
where first term is the energy expended for elastic deformation, and second term is the
energy of primary deformation.

The transformation of macromolecule conformational state takes place due to the
action of external force. Within the used approach, the force acts on external
component of macromolecule deformation, and then, due to the coupling of external and
internal components affects the macromolecule conformational state. From the results,
obtained in Sec. \ref{sec3}, it is seen, that modification of macromolecule conformational
state is going through the changes of parameters $\epsilon_{\chi}$ and ${\ae}$, which
values are determined by the magnitudes of $\epsilon$ and $k_R$, and by parameter of
components coupling $\chi$. If parameters $\epsilon$ and $k_R$ are the constants for
definite macromolecules, the parameter $\chi$ has to depend on acting force, and to
change like a magnitude of external force.

Let us consider the dependencies of $\epsilon_{\chi}$ and ${\ae}$ on $\chi$ .
It is obviously, that if $f$=0, the definite coupling between macromolecule transformation
components exists, let it is $\chi_o$. If $f$$\neq$0, the force can affect the coupling
between the components, as is supposed $\chi\propto{f}$. Accordingly to
(\ref{eq:eps},\ref{eq:epsK},\ref{eq:kap}), value of $\chi$ can change in definite numerical
interval, that is: $\chi_o<\chi<\chi_{lim}$, where  $\chi_{lim}=\sqrt{\epsilon k_R}$.
The change of coupling parameter leads to the changes of $\epsilon_{\chi}$ and ${\ae}$, and
to shaping of macromolecule conformation. When $\chi\rightarrow\chi_o$, then
${\ae}\rightarrow{\ae}_o$ and $\epsilon_{\chi}\rightarrow\epsilon_{\chi o}\sim\epsilon$. When
$\chi\rightarrow\chi_{lim}$, then ${\ae}\rightarrow\infty$ and $\epsilon_{\chi}\rightarrow0$.

\begin{figure}[h!]
\begin{center}
\includegraphics[width=7.0cm]{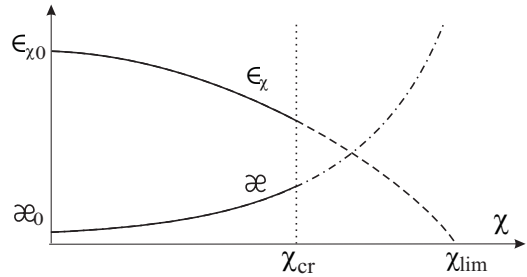}
\caption{The dependencies of parameters ${\ae}$ and $\epsilon_{\chi}$ from value of
coupling parameter $\chi$. In the figure a critical value of coupling parameter
is indicated.}
\label{kappa}
\end{center}
\end{figure}

The numerical analysis of expressions (\ref{eq:eps},\ref{eq:kap}) with possible values of
$\epsilon$ and $k_R$ shows the validity the conclusions drawn (see Fig. \ref{kappa}).
From the results presented in the figure, it is seen that parameter ${\ae}$ grows with
peaking under increasing of $\chi$. Under the critical value of acting force, coupling
parameter reaches the magnitude $\chi_{cr}$. In its turn, the parameter of primary deformation
achieves its critical value ${\ae}_{cr}$, which accords to appearance of conformational
bistability in macromolecule structure. In opposite, parameter ${\epsilon_{\chi}}$
decreases with increasing of $\chi$. Such a course of the dependence explains the
decreasing in the value of conformational barrier in macromolecule under tension of
external force, as shown in Fig. \ref{Crit-energy}. Once the coupling parameter
riches its peaking value, the further changes in macromolecule characteristics are
terminated, and another stage of threshold deformation begins.

It is useful to determine the area of plausible values of $\chi_{cr}$. Really, when parameter
${\ae}$ accords to it critical value, then from expression (\ref{eq:kap}) should be:
\bea \chi^{2}_{cr} = \frac{\ae_{cr}}{1+\ae_{cr}}\epsilon k_R \, . \label{eq:Chi}\eea
Thus, for example, if observed value of ${\ae}_{cr}=1$, then following estimation can be
done: $\chi_{cr}=\sqrt{\epsilon k_R/2}$.

It is of interest to express the value of critical force in terms of the
macromolecule characteristics. Using expression (\ref{eq:Fo}) and equality
(\ref{eq:e11}) one can obtain the value of external force in conditions of critical
regime:
\bea f_{cr}= \frac{\chi_{cr}}{{\ae}_{cr}} F(r_{2f})
 \, . \label{eq:f} \eea
The relationship of critical force magnitude with macromolecule parameters will be
better seen after rewriting expression (\ref{eq:f}) for the square of critical force.
Taking into account (\ref{eq:Chi}), expression (\ref{eq:f}) can be
presented as:
\bea f^{2}_{cr}=
 \frac{2\,\varepsilon_{om}\,k_R}{{\ae}_{cr}(1+{\ae}_{cr})}\, . \label{eq:f2} \eea
Here $\varepsilon_{om}$ is the energy difference between the ground and metastable
(stretched) state of macromolecule monomer link itself, in the absence of external force:
\bea \varepsilon_{om}=\frac{1}{2}\epsilon\,F^2(r_2)\, . \label{eq:er2} \eea
Under writing expressions (\ref{eq:f2},\ref{eq:er2}), it is taken into account, that in the
frame of accepted approximation for potential functions it is true: $r_{2f}=r_2$.

So, the square of critical force is proportional to the energy value of monomer metastable
state, and to the macromolecule stretch stiffness. As seen, the balance of product of
$\varepsilon_{om}$ and $\,k_R$ with acting force can be achieved for account of the
changing in parameter ${\ae}$ . In fact, the growth of ${\ae}$ to it peaking value is fixed
by the magnitude of critical force.

Note, as approaching of acting force to its critical value the contributions of primary and elastic
deformation increase. Wherein the elastic deformation increases linearly, the increasing of
primary deformation looks like exponential dependence. Both these contributions form the uniform
deformation of macromolecule ($\rho_{of}=\rho_{el}+\rho_{o\chi}$), but not constitute the
threshold process. However, under action of  critical force the macromolecule conformation
becomes bistable, and the conformational transition from ground to stretched state can occur in
macromolecule. Due to appearance of structural transitions, the macromolecule deformation
acquires a threshold character, and this can explain the mechanism of DNA chain overstretching.

\section{Dynamics of conformationally induced deformation in DNA macromolecule}\label{sec5}

Let us consider the occurrence  of conformational transitions and induced by them deformations
in polymorphic macromolecule of DNA type. In addressing the issue, it is necessary keep into
account, that such deformation has a local character and occurs there, where the conformational
transition occurs. The realization of structural transitions in macromolecule chain occurs
through the formation of domains with another structural states \cite{Grosberg89}. So, the
conformationally induced deformation is not uniform, and cannot cover whole macromolecule
at once. If  energies of different states are the same, the boundaries of domains (domain
walls) can move along the chain increasing (or decreasing) the domain area  \cite{Krumhansl75,BruceCowley81,Davydov88}.

Of course, the domains with stretched form can appear also in conditions of inequality of
the energies of ground and stretched states, at $f<f_{cr}$. However, due to the energy difference,
these domains will not be stable and can not take part in the process of DNA overstretching. The
probability of the reverse transition to ground state will be higher, than the for the direct process.
At the same time, the appearance of conformational bistability under $f=f_{cr}$ promotes the
occurrence of stable domains, provides the facility for conformationally induced deformation
to propagate along the macromolecule chain, and implements threshold character of deformation.

In this section the dynamics of conformational transition and induced by it deformation in DNA
macromolecule are studied for the conditions of critical regime of external force action.
Let us go to the continuum approximation, which is usually used for the studying structural
transitions dynamics. In this approximation: $R=R(z,t)$ and $r=r(z,t)$,
$(R_n-R_{n-1})\rightarrow hR'$, $(r_n-r_{n-1})\rightarrow hr'$, $R'$ and $r'$ are derivatives
on $z$, and $h$ is the distance between monomers along the macromolecular chain.

The  equations of motion for external and internal components of
conformationally induced deformation can be written as:
\bea
\ddot{R} & = & s^{2}_{R}R'' - \frac{\chi h}{M}\frac{dF}{dr}r' \, \label{eq:Rmot};
\\
\ddot{r} & = & s^{2}_{r}r'' - \frac {1}{m}\left[ \epsilon F(r)-
\chi h R'\right] \frac{dF}{dr}\, . \label{eq:rmot} \eea
In equations (\ref{eq:Rmot},\ref{eq:rmot}) $s^{2}_{R}=k_{R}h^{2}/M$, $s^{2}_{r}=k_{r}h^{2}/ m $,
and the relation between potential functions ($\Phi(r)$ and $F(r)$) is used also.

To consider the dynamics of conformational transition along macromolecular
chain, let us introduce a wave coordinate: $\zeta=z-vt$, and look for the solution
having asymptotic of ground ($r_{0f}$) or metastable ($r_{2f}$) states of macromolecule.

After wave substitution and one-time integration of equations (\ref{eq:Rmot}),
one can obtain the following:
\be
(s^{2}_{R}-v^2)R_{\zeta} - \frac{\chi h}{M}\, F(r) = C_R \, .
\label{eq:Rzmot}
\ee
Here $R_{\zeta}$ is derivation on $\zeta$, $C_R$ is the constant of integration.
Accordingly to accepted initial conditions $C_R= fh/M$.
Equation (\ref{eq:Rzmot}) determines the dynamics of deformation in macromolecule.

Substituting  expression for $R_{\zeta}$ from equation (\ref{eq:Rzmot}) in equation
(\ref{eq:rmot}), and integrating it, the equation for conformational component can be obtained:
\be r_{\zeta}^{2}+ Q^{2}(r) = 0\, , \label{eq:rZ}\ee
where:
\be \begin{split}
Q^{2}(r)= \frac{1}{m (s^{2}_{r}-v^2)}\biggl\{ - \epsilon F^{2}(r)- \frac{{\ae}}{k_R} f^{2}+ \\
+ \frac{\chi h^2}{M (s^{2}_{R}-v^2)}\biggl[\chi F^{2}(r) +2 f F(r)\biggl]\biggl\} \, .
\end{split} \label{eq:Qr} \ee
Under writing expression (\ref{eq:Qr}), it is taken into account that the constant of integration
accords to expression (\ref{eq:Cr}).

Unfortunately, equation (\ref{eq:rZ}) with potential (\ref{eq:Qr}) is quite complicated for direct
analysis. But this equation can be simplified, taking into account that in experiments on DNA
threshold deformations the measured processes have the velocities not exceeding $10^{-5}$ m/sec \cite{Cluzel+96}, which are much smaller then the velocity of sound in DNA ($s_{R}\sim10^3$ m/sec \cite{Lee-87}). The velocity $s_r$ should have close order $s_R \leq s_r$, accordingly to
analysis in \cite{TheorBiol90}. Therefore, let us consider the problem in approximation of a
small values of transition wave velocity: $v^{2}\ll s^{2}_R$,$s^{2}_r$. In this case the
potential function (\ref{eq:Qr}) takes the form which accords to energy (\ref{eq:E-r}) with
constant (\ref{eq:Cr}), and at $f=f_{cr}$ to expression (\ref{eq:Ecr}), so that
\be Q^{2}(r) = - \frac{\epsilon_{\chi}}{k_{r}h^{2}} \, \mathcal{F}^{2}_{cr}(r) \,.
\label{eq:Qzeta} \ee

The solution of equation (\ref{eq:rZ}) with potential (\ref{eq:Qzeta}) for asymptotic
of stable states (at $\zeta\rightarrow \pm \infty$, $r \rightarrow r_{0f}$ or $r_{2f}$,
and $r_{\zeta}\rightarrow0$) has the form of domain wall \cite{Krumhansl75,BruceCowley81,Davydov88}.
After integration of equation (\ref{eq:rZ}) with mentioned above boundary conditions,
one can obtain the expression for the domain wall (conformational transition wave)
in DNA macromolecule:
\be r(\zeta) = r_{0f} + d \, [1 \pm {\rm th}(q_{r} \zeta)] \,, \label{eq:rzeta} \ee
where constant
\be q_{r} = \sqrt{\epsilon_{\chi}/k_{r}d^{2} h^{2}}  \, \label{eq:qr} \ee
has the dimension of inverse length.

In expression  (\ref{eq:rzeta}) the sign has been chosen in accordance to the boundary
conditions: plus, if at the boundary $r_{0f}$ state is realized; minus, if $r_{2f}$.
As one can see, solution (\ref{eq:rzeta}) is a wave in the form of a step - the transition
from state $r_{0f}$ to $r_{2f}$, or reverse process.

Taking into account result (\ref{eq:rzeta}), let us find the deformation induced in
macromolecule by the appearance of conformational transition. The equation for external
component (\ref{eq:Rzmot}) in the approximation of small velocities takes the form:
\be
R_{\zeta} = \frac{1}{h k_R} \left[ f_{cr} + \chi F(r_{0f}) + \chi \mathcal{F}_{cr}(r) \right]
 \, , \label{eq:Rzeta}\ee
where formula (\ref{eq:E-rF}) is used as well.

\begin{figure}[h!]
\begin{center}
\includegraphics[width=8.6cm]{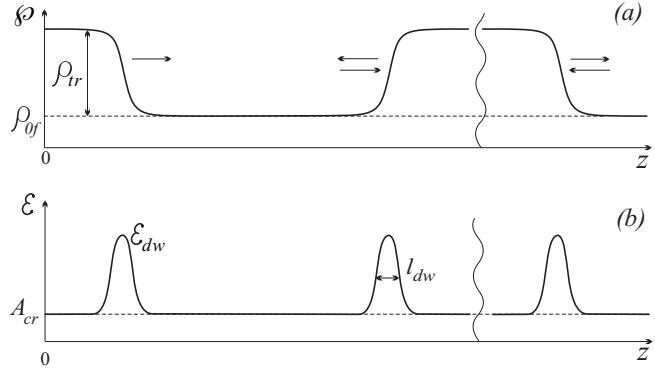}
\caption{Propagation of the wave of conformation deformation along macromolecule chain in the
conditions of critical regime of external force action: (a) macromolecule deformation comprises
the deformation induced by conformational transition ($\rho_{tr}$), and the sum of elastic and
primary deformations of macromolecule structure ($\rho_{of}$); (b) energy distribution
in macromolecule chain under threshold deformation, $\varepsilon_{dw}$ and $l_{dw}$ is energy
and thickness of domain wall, $\mathcal{A}_{cr}$ - the work of external force to form bistability
in macromolecule.}
\label{solitony}
\end{center}
\end{figure}

Accordingly, the deformation of monomer link of polymorphic macromolecule can be written as:
\be
\wp(\zeta)=h R_{\zeta}=\rho_{of} + \rho_{tr}(\zeta) \, ,
\label{eq:nRO}\ee
where $\rho_{of}=\rho_{el}+\rho_{o\chi}$ is the uniform deformation of macromolecule chain,
and $\rho_{tr}$ is the deformation of the monomer link caused by conformational transition (Fig. \ref{solitony}a).

The part of deformation induced by conformational transition is expressed as:
\be
\rho_{tr}(\zeta)= \frac{\chi}{h k_R} \int \mathcal{F}_{cr}(r) d{\zeta}=
\frac{1}{2} \rho_{tr} [1 \pm {\rm th}(q_{r} \zeta)]
\,,
\label{eq:ROzeta}\ee
where $\rho_{tr}= 2{\chi}/{hk_{R}\,q_{r}}$. Here the solution for $r(\zeta)$ (\ref{eq:rzeta}) and
expression (\ref{eq:Fcr}) for function  $\mathcal{F}_{cr}(r)$ are used also. To derive expression
(\ref{eq:ROzeta}), it is supposed, that at one side of macromolecular chain the state $r_{2f}$ is realized, and at another side - $r_{0f}$ state. Thus, $\wp = \rho_{of}+\rho_{tr}$ at one side of domain wall (where $r=r_{2f}$), and $\wp = \rho_{of}$ at another side ($r=r_{0f}$) (Fig. \ref{solitony}a).

It should be noted, that solutions \ref{eq:rzeta} and \ref{eq:ROzeta} obtained for domain
walls are the topological solitons. This type of solitins remains stable for unchanged
boundary conditions in macromolecule chain \cite{Krumhansl75,BruceCowley81,Davydov88}, and their appearance in the macromolecule happens at the end of the chain only.
The equations (\ref{eq:rZ}) and (\ref{eq:Rzeta}) have also the two-solitons solutions.
These solutions are shown in Fig. \ref{solitony} as well. Such type of solutions describes
the occurrence of the domains of another conformation inside macromolecule. Note, that
realization of single-soliton solution is more favorable energetically, because it requires
only one domain wall formation at the end of macromolecule chain in this case.

A single-soliton and two-solitons solutions differ from each other not only by the
number of domain walls, but also by the boundary conditions for their realization.
For the realization of single-soliton solution it is important to have distinct
stable states ($r_{0f}$ or $r_{2f}$) at the different ends of transition region. For
realization of two-solitons solution it is necessary to have the same stable states
(usually ground state $r_{0f}$) at the domain boundaries.

\section{Properties of overstretching transition: quantitative estimations}\label{sec6}

The developed approach allows to study the properties of overstretching process in dsDNA on
quantitative level. It is of interest to determine amplitude, thickness and energy of domain
walls for the regime of dsDNA critical deformation.

The expression for amplitude of macromolecule deformation, induced by conformational
transition, can be obtained substituting the value of constant $q_{r}$ (\ref{eq:qr})
in expression of $\rho_{tr}$ from (\ref{eq:ROzeta}). So, $\rho_{tr} = 2d \sqrt{{\ae}k_{r}/k_{R}}$.
Considering that constants $k_{r}$ and $k_{R}$ for DNA macromolecule are the same order
of values, it can be obtained a simple formula for the estimation of deformation amplitude:
\be
\rho_{tr} \approx 2d \sqrt{{\ae}} \,.
\label{eq:AR}\ee
Note, whereas the amplitude of conformational transition $d$ is some constant defined
by the pathway of transition, the amplitude of macromolecule deformation varies proportionally
with square root of ${\ae}$.

Using expressions (\ref{eq:deform},\ref{eq:def2},\ref{eq:rooChi}) and (\ref{eq:AR}), the
macromolecule deformation per one monomer unit can be written as:
\be
\wp = \rho_{el}(1+{\ae})+ 2d\sqrt{{\ae}} \,.
\label{eq:RO}\ee

The thickness of domain wall is determined as $l_{dw}\approx 2q^{-1}_{r}$ \cite{Krumhansl75}. So,
using expressions (\ref{eq:eps},\ref{eq:kap}) and (\ref{eq:qr}), it can be obtained the following
formula for thickness of domain wall under macromolecule overstretching:
\be  l_{dw} \approx  h\,  \frac{k_R \,\rho_{tr}}{\chi} \,.
\label{eq:ld}\ee
It is seen, that domain wall thickness (\ref{eq:ld}), as well as its amplitude (\ref{eq:AR}),
is proportional to root of ${\ae}$.

The energy of domain wall can be determined as:
\be  \mathcal{E}_{dw}= \frac{1}{2h} \int_{-l_{dw}/2}^{l_{dw}/2} \mathcal{E}_{cr}(\zeta)d{\zeta} \,,
\label{eq:Ed}\ee
where
\be  \mathcal{E}_{cr}(\zeta) =  \epsilon_{b}\,\, ch^{-4}(q_{r}\zeta) \,.
\label{eq:Ecrd}\ee
To derive expression (\ref{eq:Ecrd}), the solution (\ref{eq:rzeta}) and expressions
(\ref{eq:Ecr},\ref{eq:Echi}) are used.

After integration in (\ref{eq:Ed}) with expression (\ref{eq:Ecrd}), the resulting formula
has the view:
\be  \mathcal{E}_{dw}= 0.61 \,\frac{l_{dw}}{h} \epsilon_{b} \,.
\label{eq:Edw}\ee

To estimate the full energy required to initiate domain wall formation,
it is necessary to include in consideration the work of external
force to create the bistability in macromolecule chain (\ref{eq:Acr}).
So, the initiation energy of domain wall formation can be written as:
\be  \mathcal{E}_{id} = \mathcal{E}_{dw}+ \mathcal{A}_{dw}\,,
\label{eq:Ein}\ee
Here $ \mathcal{A}_{dw}$=$\mathcal{A}_{cr}\,l_{dw}/{h}$.
In expression (\ref{eq:Ein}) the first part of energy is connected with the domain
formation as such. Another term accords to the part of work on creation of
conformation bistability on the interval of domain wall thickness.

The obtained expressions are used for calculations of the characteristics of dsDNA
overstretching process: the contributions from elastic, primary and threshold
deformations, the energies and sizes of domain walls, the values of energy barrier and
energy difference between ground and stretched states under overstretching
transition in macromolecule. The results of calculations are shown in Tabl.
\ref{tab:1}.

In calculations the following magnitudes of model parameters are
used: $h$=3.4 ${\AA}$; $k_R$=S$/h$, where dsDNA stretch modulus S=1100$\pm$200 pN
(accordingly to \cite{Gore06}). That is $k_R$=4.6$\pm$0.8 kcal/mol ${\AA}^2$. The
value of parameter $\epsilon$=4 kcal/mol is adopted by the data of conformational
analysis of DNA backbone \cite{Olson82,Saenger}. This choice accords to the assumption of
\cite{SSmith+B96,KonradBlonick96} about the main role of conformational transformations
in DNA sugar rings during the overstretching transition. Parameter
$d$ is taken to be equal to 1 $\AA$. That value by order of magnitude corresponds
to the data of conformational analysis of overstretching transformation in dsDNA
\cite{LebrunLavery96} (stucture 1.6), and \cite{KonradBlonick96}.

For estimations of other characteristics of dsDNA overstretching it is
necessary to know the value of parameter $\ae$ (or $\chi$). The estimations of these
parameters can be done from observed value of macromolecule deformation in the critical
regime. As known \cite{SSmith+B92,Cluzel+96,SSmith+B96,Bustamante00,Lavery02,Bustamante03}
for critical regime, when $f_{cr}$=65 pN, the deformation of dsDNA with average nucleotide
content comprises 0.7 of its length. Considering, that the stretch modulus of such dsDNA has
average value (S=1100 pN), from expression (\ref{eq:RO}) it can be calculated
$\ae_{cr}\approx$1. This estimation is important for the evaluation of coupling parameter
$\chi$. Using formula (\ref{eq:Chi}) with known values of $\epsilon$ and $k_R$, it can be
obtained $\chi_{cr}\approx$3 kcal/mol$\AA$.

It should be emphasized, that estimated in such way parameters accord to the stretching
of DNA chain with average value of stretch modulus. But dsDNA is a  heteronomous
macromolecule, and its chain consists from the fragments with different nucleotide content
(and so with distinct elastic properties). So, it is necessary to consider, that under
action of definite critical force the fragments with other (not average) elastic
properties are stretched also. Tacking this into account, the calculations of double
helix elongation characteristics are performed also for DNA fragments with stretch
modules differing from their average value (on $\pm$ 200 pN). These calculations are
done with the same values of parameters $\epsilon$ and $\chi$ for all macromolecule
fragments. This is because parameter $\epsilon$ reflects the conformational transformations
in dsDNA backbone and has no change considerably for different nucleotide content. In its turn,
coupling parameter $\chi$, as assumed, is depended on the value of external force, but the
force is the same for whole macromolecule. However, the values of parameter $\ae$ for these
distinct DNA fragments are calculated by formula (\ref{eq:kap}) with distinct (not average)
magnitudes of $k_R$.

It is of interest also, to estimate the values of energy barrier for macromolecule transition
to stretched state, and energy difference between ground and stretched states under
overstretching transition for DNA fragments with distinct stretch modules. The value of
energy barrier can be calculated as $\epsilon_{b}$=$\mathcal{E}(r_{1f})$, using expressions (\ref{eq:eps},\ref{eq:Echi}) and already determined parameters. The energy difference between the
states in the frame of developed approach is determined as $\epsilon_{m}$=$\mathcal{E}(r_{2f})$
accordingly to (\ref{eq:Me}) (note, $\mathcal{E}(r_{of})$=0).
Needed for these calculations value of $F(r_{2f})$ is equal to the value of $F(r_{of})$
in critical regime and can be found from expression (\ref{eq:Fo}). For the fragments with
distinct stretch modules value of $F(r_{2f})$ can be calculated with the same values of
$f_{cr}$ and $\chi_{cr}$, but with value of $\ae_{cr}$ calculated by (\ref{eq:kap})
for different stretch modules. As seen, the values of ($\epsilon_{b}$) and ($\epsilon_{m}$)
can be calculated as well.

The results of calculations of overstretching characteristics for dsDNA per one monomer
unit for distinct stretch modules but with the same value of external force ($f_{cr}$=65 pN)
are shown in  Tabl. \ref{tab:1}. The data for critical regime realization (for DNA with
average stretch modulus) are presented in the second row. All values in the table are
calculated with the accuracy to second sing after coma and then are rounded to first sing.

\begin{table}
\begin{center}
\caption{The values of parameter $\ae$, stretching deformations of DNA monomer link, energies
of transition barrier and energy differences between the states,  thickness of domain
walls and energies of their formation in dsDNA under tension of critical force ($f_{cr}$=65 pN)
for heteronomous macromolecule with average stretch modulus (S=1100 pN) and for DNA fragments
with distinct (on $\pm$200 pN) stretch modules.}
\label{tab:1}
\begin{tabular}{c|c|c|c|c|c}
\hline
S\,&\,$\ae$\,&\,$\wp$\,\,($\rho_{el};\,\rho_{o\chi};\,\rho_{tr}$)\,
&\,$\epsilon_{b}$\,($\epsilon_{m}$)\,&\,$l_{dw}$\,
&\,$\mathcal{E}_{id}$\,($\mathcal{A}_{dw};\mathcal{E}_{dw}$)\, \cr
(pN)\,&\,$(-)$\,&\,$(\AA)$\,&\,$(\frac{kcal}{mol})$\,
&\,$(h)$\,&\,$(\frac{kcal}{mol})$\, \cr
\hline
&&&&& \cr
900\,&\,1.5\,&\,3.0 \,(0.2; 0.4; 2.4)\,&\,0.8(-0.1)\,&\,3.1\,&\,2.4\,(0.9; 1.5)\, \cr
1100\,&\,1.0\,&\,2.3 \,(0.2; 0.2; 1.9)\,&\,1.0\,\,(0.0)\,&\,3.0\,&\,2.4\,(0.5; 1.9)\, \cr
1300\,&\,0.7\,&\,2.0 \,(0.2; 0.1; 1.7)\,&\,1.2\,\,(0.1)\,&\,3.1\,&\,2.6\,(0.4; 2.2)\, \cr
\end{tabular}
\end{center}
\end{table}

\section{Discussion: theory and experiment}\label{sec7}

\subsection{Quantitative agreement}

As seen, the calculated values of dsDNA overstretching deformation (Tabl. \ref{tab:1}) in
order of their magnitudes correspond to experimental data
\cite{SSmith+B92,Cluzel+96,SSmith+B96,Bustamante00,Lavery02,Bustamante03}. Over the range
of DNA stretch modulus variation the contribution of elastic deformation $\rho_{el}$ to
total value of macromolecule deformation changes moderately, only in second sign after
coma. In contrast, the contributions of primary and threshold deformations and the value
of parameter ${\ae}$ change significantly, and are larger for macromolecule fragments with
smaller stretch modulus. The deformation $\rho_{tr}$ gives the largest
contribution to the value of macromolecule deformations in each case. The large
values of $\rho_{tr}$ provide the threshold character of macromolecule deformation
as a whole. The obtained result accords with the data of overstretching kinetics study
in force-clamp experiments \cite{FuYan11,Bianco11}. As shown in these works, the amplitudes
of double helix elongation increase in times in the interval of critical force
action, where DNA macromolecule exhibits threshold deformation.

The results presented in Tabl. \ref{tab:1} also show, that the transition barrier between
compact and stretched DNA forms in conditions of critical regime has the order of 1 kcal/mol.
The energy difference between transition states is absent for DNA with the average stretch
modulus (as expected for critical regime), and is sufficiently small (but not zero) for DNA
fragments with distinct stretch modules. It means, that propagation of domain walls in
heteronomous macromolecule occurs with minimal losses just for the conditions of critical
regime. With increasing the deviation from these conditions, the losses will increase, and the
deformation will not be pass as threshold process, and not have large contribution in whole
macromolecule deformation. This result explain the relatively small width of the force interval
of critical regime realization in dsDNA, observed in experiments \cite{SSmith+B92,SSmith+B96,Bustamante00,Lavery02,Bustamante03}.

The calculated energies of domain walls initiation (Tabl. \ref{tab:1}) are more than twice
the energies of transition barriers, that should be facilitated domain walls passage through
the barriers. The domain walls energies do not change significantly for macromolecule fragments
with distinct stretch modules. In such an effect the simultaneous decreasing of the work
$\mathcal{A}_{cr}$ and increasing of domain wall energy play its role. Similar effect is
seen for the domain walls sizes ($l_{dw}$), which values are conserved on interval
of stretch modulus variation. Here two processes, the decreasing of parameter ${\ae}$ and
increasing of macromolecule stiffness, proceed concurrently (\ref{eq:AR},\ref{eq:ld}). Thus,
the domain walls features are weakly dependent on variations of stretch modulus, under its
considered deviations from average value ($\pm$ 200 pN).

From calculated data it is seen also, that the energy of domain wall formation ($\mathcal{E}_{dw}$)
introduces the larger contributions to whole initiation energies of overstretching transition.
It is of interest to compare the obtained values of $\mathcal{E}_{dw}$ with the data of
thermodynamical analysis of DNA overstretching transitions in \cite{ZhangYan12}. In this work the
free energy difference for overstretched transitions are measured for definite DNA sequences, where
are observed $B$-$S$ or FIM transitions. By the conditions of experiment the transitions are
studied in the critical force interval (\cite{ZhangYan12}). From these data the values of free
energy for dsDNA overstretching transitions can be calculated by known formula $\triangle G$=
$\triangle H$-$\triangle ST$. The calculations give the following values: 1.7 kcal/mol for DNA
overstretching in FIM transition, and 1.9 kcal/mol for overstretched DNA resulting in $B$-$S$
transition. As seen, these energy values agree with the results listed in Tabl. \ref{tab:1} for
the domain wall energy $\mathcal{E}_{dw}$ in DNAs with less and more helix stiffness (first and
second rows).

The obtained magnitudes of domain walls energies allow also to estimate the cooperativity
of $B$-$S$ transition in conditions of critical regime. Using formula for cooperativity length \cite{Grosberg89}, which in this case has a view: $\nu_{tr}= exp(\mathcal{E}_{dw}/kT)$, and
the data on domain energy for the conditions of critical regime (Tabl. \ref{tab:1}, second row)
one can obtain the length of cooperativity (average domain size) for DNA $B$-$S$ transition
equal to about 24 base pairs. This estimation fully accords to the value obtained in \cite{Bianco11}
from the analysis of overstretching transition kinetics in dsDNA.

Of course, performed calculations are the estimations only. But the data of Tabl. \ref{tab:1}
show, that proposed theoretical approach can be used for the interpretation of experimental data.

\subsection{Dependence on nucleicleotide content. $B$-$S$ and FIM overstretching transitions}

Up to now in the present study, when considering the process of DNA overstretching, no differences
between two scenarios of double helix elongation, $B$-$S$ or FIM transitions, have been done. Now,
on the basis of obtained results, the realization of two possible processes in heteronomous dsDNA
can be analyzed.

From obtained results (Tabl. \ref{tab:1}) it is seen, that for double helix fragments with
different values of elastic and conformational properties the overstretching deformation should
be different. As one can estimate from results of \cite{Rief99,Gaub2000}, the stretch modulus
for A$\cdot$T-rich DNA is smaller than for G$\cdot$C-rich DNA. So, accordingly to the estimations
made (Tabl. \ref{tab:1}), and expression (\ref{eq:RO}), the overstretching deformation of
heteronomous DNA should be larger for macromolecule fragments with smaller stiffness, that is,
for A$\cdot$T-rich fragments. Note, this conclusion is in accordance with results of \cite{Norden12},
where the larger amplitudes of overstretching deformation are observed for A$\cdot$T-rich DNA.

In its turn, the larger elongation of some dsDNA fragments in overstretching process has to lead
to larger loss in base pairs stacking for the double helix. Such assumption follows from the
consideration of possible structures for overstretched dsDNA. Accordingly to conformational
analysis \cite{LebrunLavery96,KonradBlonick96}, in overstretched double helix the base pairs
are greatly inclined with large slide and shift.  All these structural changes should increase
with the growth of double helix elongation, and should lead to losses in base pairs stacking.
As known \cite{MFK-06}, the base stacking on a par with the hydrogen bonding determines the
base pairing in the double helix, and the stability of dsDNA as a whole. Thus, the large
overstretching deformation of DNA fragments more likely to be lead to base pairs disruption
and subsequent internal melting. Can assume, that this is valid for the overstretching of
DNA fragments with lower stretch modulus, such as A$\cdot$T-rich sequences.

It can be assumed, that $S$-form is stable for dsDNA with sufficiently large content of
G$\cdot$C pairs, and less (or not) stable for A$\cdot$T-rich DNA. Such assumption agrees with the
results of experiment \cite{Norden12}, where under overstretching of A$\cdot$T-rich oligonucleotides
the intermediate (between $B$ and melted) state is observed. As shown in \cite{Norden12}, this
intermediate state is not stable, and transmits to melted state. It can be just the observation
of metastable $S$-form in A$\cdot$T-rich DNA. Hence, under overstretching of DNA with rich content
of A$\cdot$T pairs, or dsDNA with reduced stability (temperature increasing, or counterions
concentration decreasing), the overstretching process should lead to force induced melting transition.

\subsection{Ability of $B$-$S$ transition propagate along DNA chain}

Important point of proposed mechanism of DNA overstretching is the ability of the domain boundaries
to move along the double helix. Whether domain walls will propagate through macromolecule or not
depends on the elastic properties of DNA chain, its homogeneity, and the external environment (which
can slow down or stop the moving). The necessary conditions of domain walls movement in DNA are the
preservation of elasticity of the macromolecule chains and relatively low transition barriers between
$B$ and $S$ conformations (accordingly to known conditions for domain walls propagation \cite{Krumhansl75}).

It is known, from the results of structure organization study of dsDNA \cite{Saenger}, the elastic
stiffness of DNA macromolecule is mainly caused by the stacking interactions between DNA base pairs
inside the helix. As evidenced the results of conformational analysis of dsDNA $S$-form
\cite{LebrunLavery96,KonradBlonick96} and observed lack of hysteresis for $B$-$S$ transition in
dsDNA \cite{FuMarcoYan10,FuYan11}, under formation of $S$ structure the stacking is largely
retained. Saving the base pairs stacking in the case of $B$-$S$ transition provides the sufficient
grounds and evidence for preservation of elastic properties of macromolecules. Together
with the lowering of potential barriers between $B$ and $S$ states in the critical regime (Fig.
\ref{kappa}), the stacking  preservation allows to consider, that domain walls movement in $B$-$S$
transitions is possible. That is not so for FIM transition, because under this transition the elastic
properties of DNA chain are not retained, due to transition of the double helix to no ordered state.

It is important, that the energy losses for propagation of $B$-$S$ transition are considerably less,
than for FIM.  The losses in $B$-$S$ transition for internal component is minimal, because in
this case the base pairs move as a unit, which masses are the same. Besides, under $B$-$S$ transition
the losses for external component from the interaction with surroundings cannot be large also. This
is so, because under $B$-$S$ transition the double helix diameter is significantly reduced (\cite{LebrunLavery96,Maaloum11}). In this sense, the transition from $B$ to $S$ form is advantageous.
The ability of $B$-$S$ transition for propagation in heteronomous DNA macromolecule is supported also
by the results of quantitative estimations shown in Tabl. \ref{tab:1}. As seen, the characteristics
of domain walls moving along heteronomous macromolecule can remain sufficiently stable under
variability of dsDNA stretch modulus. In addition, the domain wall energy exceeds in twice
the energy of transition barrier (Tabl. \ref{tab:1}). These factors should support the possibility of
domain walls motion under $B$-$S$ transition in dsDNA. Of course, the definite level of dissipation
remains for $B$-$S$ transition dynamics in heteronomous DNA (in particular, from the variations of
stacking interactions along the chain), but this is second-order effects.

Another picture of overstretching process is realized in FIM transition. Under FIM passing the
hydrogen bonds in the base pairs are becoming broken. Note, the energy of pairs ruptures are different
for A$\cdot$T and G$\cdot$C pairs. In addition, the masses of bases are different also. These factors
contribute the additional dissipation to FIM transition propagation in compare with $B$-$S$. Besides,
under FIM transition the diameter of the double helix should increase, which together with the
dissipation from the rupture of hydrogen bonds, should lead to an inhibition of possible motion
bits.

Therefore, FIM transition in DNA chain takes place as slow, nearly static, structure transformation
in comparing with fast $B$-$S$ transition. Such difference in the dynamics of overstretched transitions is observed in experimental studies \cite{FuMarcoYan10,FuYan11}, as well.

\section{Conclusions}\label{sec8}

On the basis of obtained results some features of the mechanism of threshold
elongation in dsDNA can be drawn.

As shown, the overstretching deformation of polymorphic macromolecule of DNA type
consists of following contributions: elastic deformation ($\rho_{el}$), deformation due
to primary changes in macromolecule structure before structural transition ($\rho_{o\chi}$),
and threshold elongation induced by conformational transition as such ($\rho_{tr}$). This
conclusion is in fully agreement with data \cite{SSmith+B96} (Fig.1), and all experimental
studies of DNA overstretching. The estimated values of macromolecule overstretching in the
critical regime per one monomer unit (Tabl. \ref{tab:1}) accords to experiments on $\lambda$-DNA \cite{SSmith+B96}. The greatest contribution to overstretching process (more then 80$\%$) in
dsDNA is introduced by threshold elongation. This conclusion corresponds to the experimental
results of \cite{FuYan11,Bianco11}, where the largest amplitudes for dsDNA stretching are
observed for the interval of critical force action.

According to developed theory, the overstretching process in dsDNA passes in two stages.
Firstly, at $f<f_{cr}$, the conformational restructuring in DNA double helix
takes place. The manifestations of these restructuring are the shift of ground state
position in macromolecule conformation, and the appearance and subsequent growth of
primary deformation in macromolecule chain. As shown, the structural changes are
necessary to transform macromolecule to bistable conformation, and to decrease the energy
barrier between transition states. Note, that at $f<f_{cr}$ the domains with stretched
conformation can occur in macromolecule chain, as well. But, due to the difference in the
energy of compact and stretched states, these domains will not be stable, and can not take part
in the process of DNA overstretching.

The confirmation of the double helix restructuring before overstretching transition should
be found in DNA low-frequency vibrational spectra, where the soft modes in spectra range
below 200 cm$^{-1}$ \cite{UrabeTominaga83,VolkovKosevich91} can be observed under the
double helix conformation transitions.
Note, that obtained in the present study effect of the growth with peaking of parameter
${\ae}$ (Fig. \ref{kappa}), and primary deformation ($\rho_{o\chi}$) as a whole, accords to
the observed exponential course of macromolecule elongation in the kinetics of dsDNA
overstretching process (\cite{Bianco11}). As seen from the developed theory, the
achievement of parameter ${\ae}$ its peaking value corresponds to the formation
bistability in macromolecule conformation.

On the second stage (at $f$=$f_{cr}$), the bistability in the double helix structure formed
with simultaneously reducing of transition barrier between compact and stretched conformations.
In this conditions the domains with stretched conformation can occur in DNA chain with large
probability. In the critical regime domain walls get the ability to propagate along DNA
macromolecule, to increase the domains sizes, and to spread the overstretching transition.
By estimation made, the length of cooperativity for $B$-$S$ transition in the
conditions of critical regime is about 24 base pairs, that accords to the value obtained
in \cite{Bianco11} from the analysis of DNA overstretching kinetics. Also, calculated in
the present study value of domain wall energy (1.9 kcal/mol, for average stretch modulus
in Tabl. \ref{tab:1}) corresponds to the data of thermodynamical analysis of DNA
overstretching in $B$-$S$ transition \cite{ZhangYan12}.

The proposed mechanism of DNA overstretching suggests, that the external force reaches its critical
value to produce the work on bistability occurrence in macromolecule chain. As seen from (\ref{eq:f2}),
the value of critical force is proportional to the macromolecule stiffness ($k_R$),
the magnitude of energy difference between the ground and metastable (stretched) state in
monomer link ($\varepsilon_{om}$), and inversely proportional to the parameter of primary
deformation (${\ae}$). Since the values of $\varepsilon_{om}$ and $k_R$ should be considered
as macromolecule constants, the critical balance between the acting force and macromolecule
characteristics should be obtained through the changes in parameter ${\ae}$, and, accordingly,
through the macromolecule restructuring.

As discussed in \ref{sec7}.C., the propagation of $B$-$S$ transition boundaries is very
probable in DNA structure, because of small rate of energy dissipation, as inside of the
double helix and outside the macromolecule also. Another option is the propagation of FIM
transition, where the energy dissipation should be considerably larger due to the rupture
of hydrogen bonds in the base pairs. This conclusion gives an explanation of the fact
\cite{FuMarcoYan10,FuYan11}, that $B$-$S$ transition in dsDNA is implemented as fast process,
but FIM transition - as slow, nearly static process.

In this way, the proposed mechanism of DNA deformation provides the threshold nature of the
double helix overstretching. As seen, the realization of such mechanism is possible due to
coupling between macromolecule stretching and conformation, and due to ability of macromolecule
to restructuring of its form. In this way, done in the present study assumption about the
important role of double helix polymorphism in the mechanism of dsDNA threshold deformation
allows to explain the basic features of observed effects. It should be noted, that the
absence of polymorphic properties in RNA macromolecule can be the cause of not observing for
it the overstretching effect.

\subsection*{ACKNOWLEDGMENTS}

I would like to thank Richard Lavery for sending data on conformational analysis of stretched DNA.
I also want to express my gratitude to Maxim D. Frank-Kamenetskii and Polina Kanevska for helpful discussions.

The present work was partially supported by the project of the National Academy of Sciences of Ukraine:
project 0110U007540.


\begin{thebibliography}{99}

\bibitem{Schellman74}
J.A. Schellman, Biopolymers {\bf 13}, 217 (1974).

\bibitem{BarkleyZimm79}
M.D. Barkley, B.H. Zimm, J. Chem. Phys. {\bf 70}, 97 (1979).

\bibitem{F-K83}
M.D. Frank-Kamenetskii, Comm. Mol. Cell. Biophys. {\bf 1}, 105 (1981).

\bibitem{SSmith+B92}
S.B. Smith, L. Finzi, C. Bustamante, Science {\bf 258}, 1122 (1992).

\bibitem{Cluzel+96}
P. Cluzel, A. Lebrun, C. Heller, R. Lavery, J.-L. Viovy, D. Chatenay, F. Caron, Science {\bf 271}, 792 (1996).

\bibitem{SSmith+B96} S.B. Smith, Y. Cui, C. Bustamante, Science {\bf 271}, 795 (1996).

\bibitem{Bustamante00}
C. Bustamante, S.B. Smith, J. Liphardt, D. Smith, Curr. Opin. Struct. Biol. {\bf 10}, 279 (2000).

\bibitem{Lavery02}
R. Lavery, A. Lebrun, J.-F. Allemand, D. Bensimon, V. Croquette, J. Phys.: Condens. Matter {\bf 14}, R383 (2002).

\bibitem{Bustamante03}
C. Bustamante, Z. Bryant, S.B. Smith, Nature {\bf 421}, 423 (2003).

\bibitem{Bockelmann04}
U. Bockelmann,  Curr. Opin. Struct. Biol. {\bf 14}, 368 (2004).

\bibitem{Allemand06}
J.-F. Allemand, S. Cocco, N. Douarche, G. Lia, Eur. Phys. J. E {\bf 19}, 293 (2006).

\bibitem{Prevost09}
Ch. Prevost, M. Takahashi, R. Lavery, ChemPhysChem {\bf 10}, 1399 (2009).

\bibitem{PhysLife-Williams10}
K.R. Chaurasiya, T. Paramanathan, M.J. McCauley, M.C. Williams, Phys. Life Rev. {\bf 7}, 299 (2010).

\bibitem{Killian12}
J.L. Killian, M. Li, M.Y. Sheinin, M.D. Wang,  Curr. Opin. Struct. Biol. {\bf 22}, 80 (2012).

\bibitem{Mameren09}
J. van Mameren, P. Gross, G. Farge, P. Hooijman, M. Modesti, M. Falkenberg, G.J.L. Wuite, E.J.G. Peterman,
Proc. Natl. Acad. Sci. USA {\bf 106}, 18231 (2009).

\bibitem{PaikPerkins11}
D.H. Paik, T.T. Perkins, J. Amer. Chem. Soc. {\bf 133}, 3219 (2011).

\bibitem{Rief99} M. Rief, H. Clausen-Schaumann, H. Gaub, Nature Str. Biol. {\bf 6}, 346 (1999).

\bibitem{Gaub2000} H. Clausen-Schaumann, M. Rief, C. Tolksdorf, H. Gaub, Biophys. J. {\bf 78}, 1997 (2000).

\bibitem{FuMarcoYan10}
H. Fu, H. Chen, J.F. Marko, J. Yan, Nucleic Acids Res. {\bf 38}, 5594 (2010).

\bibitem{FuYan11}
H. Fu, H. Chen, X. Zhang, Y. Qu, J.F. Marko, J. Yan, Nucleic Acids Res. {\bf 39}, 3473 (2011).

\bibitem{Norden12}
N. Bosaeus, A.H. El-Sagheer, T. Brown, S.B. Smith, B. Akerman, C. Bustamante, B. Norden,
Proc. Natl. Acad. Sci. USA {\bf 109}, 15179 (2012).

\bibitem{Liphard2001}
J. Liphardt, B. Onoa, S.B. Smith, I.Jr. Tinoco, C. Bustamante, Science {\bf 292}, 733 (2001).

\bibitem{Rouzina2001a}
I. Rouzina, V.A. Bloomfield, Biophys. J. {\bf 80}, 882 (2001).

\bibitem{Rouzina2001b}
I. Rouzina, V.A. Bloomfield, Biophys. J. {\bf 80}, 894 (2001).

\bibitem{Wuite11}
P. Gross, N. Laurens, L.B. Oddershede, U. Bockelmann, E.J.G. Peterman, G.J.L. Wuite,
Nature Physics  {\bf 7}, 731 (2011).

\bibitem{Danilovich09}
C. Danilowicz, C. Limouse, K. Hatch, A. Conover, V.W. Coljee, N. Kleckner, M. Prentiss,
Proc. Natl. Acad. Sci. USA {\bf 106}, 13196 (2009).

\bibitem{Maaloum11}
M. Maaloum, A-F. Beker, P. Muller, Phys. Rev. E {\bf 83}, 031903 (2011).

\bibitem{Bianco11}
P. Bianco, L. Bongini, L. Melli, M. Dolfi, V. Lombardi, Biophys. J. {\bf 101}, 866 (2011).

\bibitem{CoccoMarco04} S. Cocco, J. Yan, J.F. Leger, D. Chatenay, J.F. Marko,
Phys. Rev. E {\bf 70}, 011910 (2004).

\bibitem{Whitelam08} St. Whitelam, S. Pronk, Ph.L. Geissler, Biophysical J. {\bf 94}, 2452 (2008).

\bibitem{ZhangYan12}
X. Zhang, H. Chen, H. Fu, P. Doyle, J. Yan, Proc. Natl. Acad. Sci. USA {\bf 109}, 8103 (2012).

\bibitem{LebrunLavery96}
A. Lebran, R. Lavery, Nucl. Acids Res. {\bf 24}, 2260 (1996).

\bibitem{KonradBlonick96}
M.W. Konrad, J.I. Bolonick, J. Am. Chem. Soc. {\bf 118}, 10989 (1996).

\bibitem{Lauton05}
S.A. Harris, Z.A. Sands, Ch.A. Laughton, Biophys. J. {\bf 88}, 1684 (2005).

\bibitem{Marko97} J.F. Marko, Europhys. Lett. {\bf 38}, 183 (1997).

\bibitem{Gore06}
J. Gore, Z. Bryant, M. Nollmann, M.U. Le, N.R. Cozzarelli, C. Bustamante, Nature {\bf 422}, 836 (2006).

\bibitem{StormNelson}
C. Storm, P.C. Nelson, Phys. Rev. E {\bf 67}, 051906 (2003).

\bibitem{Saenger} W. Saenger, {\it Principles of Nucleic Acid Structure},
Springer-Verlag, New York, ch.9, 1984.

\bibitem{Blackburn06} G.M. Blackburn, M.J. Gait, D. Loakes, D.M. Williams (eds.) {\it Nucleic Acids
in Chemistry and Biology}, RSC Publisher, Cambridge, 383-422, 2006.

\bibitem{Ivanov94} V.I. Ivanov, L.E. Minchenkova, Mol. Biol. {\bf 28}, 1258 (1994).

\bibitem{Crothers} D.M. Crothers, T.E. Haran, J.G. Nadeau, J. Biol. Chem.  {\bf 265}, 7093 (1990).

\bibitem{Olson00} X.-J. Lu, Z. Shakked, W.K. Olson, J. Mol. Biol. {\bf 300}, 819 (2000).

\bibitem{Dickerson01} R.E. Dickerson, H.-L. Ng, Proc. Natl. Acad. Sci. USA {\bf 98}, 6986 (2001).

\bibitem{TheorBiol90} S.N. Volkov, J. Theor. Biol. {\bf 143}, 485 (1990).

\bibitem{BiophBull03} S.N. Volkov, Biophys. Bulletin (Kharkiv) {\bf 7}(2), 7 (2000);
{\bf 12}(1), 5 (2003); arXiv/q-bio/BM0312034.

\bibitem{BiolPhys05} S.N. Volkov, J. Biol. Phys. {\bf 31}, 323 (2005).

\bibitem{VolkovKanevska06} P.P. Kanevska, S.N. Volkov, Ukr. J. Phys. {\bf 51}, 1001 (2006).

\bibitem{VolkovSolov'ov07} S.N. Volkov, A.V. Solov'yov, Biophys. Bulletin (Kharkiv)
{\bf 19}(2), 5 (2007)

\bibitem{VolkovSolov'ov09} S.N. Volkov, A.V. Solov'yov, Eur. Phys. J. D {\bf 54}, 657 (2009).

\bibitem{Grosberg89} A.Yu. Grosberg, A.R. Hohlov, {\it Statistical Physics of Macromolecules},
Nauka, Moskva, 1989.

\bibitem{Krumhansl75}
J.A. Krumhansl, J.R. Schrieffer, Phys. Rev. B {\bf 11}, 3538 (1975).

\bibitem{BruceCowley81} A.D. Bruce, R.A. Cowley,
{\it Structural Phase Transitions}, Taylor and Francis Ltd., London, 1981.

\bibitem{Davydov88} A.S. Davydov, {\it Solitons in Molecular Systems}, Naukova Dumka, Kiev, 1988.

\bibitem{Lee-87} S.A. Lee, S.A. Lindsay, J.W. Powell, T. Weidlich, N.J. Tao, G.D. Lewen,
Biopolymers {\bf 26}, 1637 (1987).

\bibitem{Olson82} W.K. Olson, J.L. Sussman, J. Amer. Chem. Soc.  {\bf 104}, 270, 278 (1982).

\bibitem{MFK-06} P. Yakovchuk, E. Protozanova, M.D. Frank-Kamenetskii,  Nucleic Acids Res. {\bf 34}, 564 (2006).

\bibitem{Williams01}
M.C. Williams, J.R. Wernner, I. Rouzina, V. Bloomfield, Biophys. J. {\bf 80}, 1932 (2001).

\bibitem{Wenner02}
J.R. Wernner, M.C. Williams, I. Rouzina, V. Bloomfield, Biophys. J. {\bf 82}, 3160 (2002).

\bibitem{Mao05} H. Mao, J.R. Arias-Gonzalez, S.B. Smith, I. Tinoco Jr., C. Bustamante, Biophys. J. {\bf 89}, 1308 (2005).

\bibitem{UrabeTominaga83} H. Urabe, Y. Tominaga, K. Kubota, J. Chem. Phys. {\bf 78}, 5937 (1983).

\bibitem{VolkovKosevich91} S.N. Volkov, A.M. Kosevich, J. Biomol. Struct. Dyn. {\bf 8}, 1069 (1991).

\end{thebibliography}
\end{document}